%% file: optimise.tex
\RequirePackage[l2tabu,orthodox]{nag}
\documentclass[submission,copyright,creativecommons]{eptcs}
\usepackage{breakurl}
\providecommand{\event}{QPL 2019}

\pdfoutput=1

\input{preamble.tex}
\input{tikzpreamble.tex}

\usepackage[suppress]{rwd-drafting}

\title{Phase Gadget Synthesis for Shallow Circuits}
\author{
\begin{tabular}{ccccc}
Alexander Cowtan$^1$ & Silas Dilkes$^1$ & Ross Duncan$^{1,2,}$\thanks{ross.duncan@strath.ac.uk} & Will Simmons$^{1,}$\thanks{will.simmons@cambridgequantum.com} & Seyon Sivarajah$^1$ \\\\ 
\multicolumn{5}{c}{\footnotesize\makecell{$^1$ \textit{Cambridge Quantum Computing Ltd} \\ \textit{9a Bridge Street, Cambridge, United Kingdom}}} \\
\multicolumn{5}{c}{\footnotesize\makecell{${}^2$ \textit{Department of Computer and Information Sciences}\\
\textit{University of Strathclyde}\\
\textit{26 Richmond Street, Glasgow, United Kingdom}}} \\
\end{tabular}
}
\def\titlerunning{Phase Gadget Synthesis for Shallow Circuits}
\def\authorrunning{Cowtan, Dilkes, Duncan, Simmons, Sivarajah}

\begin{document}

\maketitle
\input{abstract.tex}
\TODO{Add additional authors}

\section{Introduction}
\label{sec:introduction}

Until fully fault-tolerant quantum computers are available, we must
live with the so-called Noisy Intermediate-Scale Quantum (NISQ)
devices and the severe restrictions which they impose on the circuits
that can be run.  Few qubits are available, but limited coherence time
and gate fidelity also limit the depth of circuits which can complete
before being overwhelmed by errors.  Automated circuit optimisation
techniques are therefore essential to extract the maximum value from
these devices, and such optimisation routines are becoming a standard
part of compilation frameworks for quantum software \cite{Nam:2018aa}.

In this paper we give an overview of some circuit optimisation methods
used in the \tket retargetable compiler platform \footnote{\tket can be 
installed as a python module via PyPI: \url{https://pypi.org/project/pytket/}}.
\tket can generate
circuits which are executable on different quantum devices, solving
the architectural constraints \cite{Alexander-Cowtan:2019aa}, and
translating to the required gate set, whilst minimising the gate count
and circuit depth.  It is compatible with many common quantum software
stacks, with current support for the Qiskit \cite{qiskit}, Cirq
\cite{cirq}, and PyQuil \cite{forest} frameworks.

Much work on circuit optimisation focuses on reducing $T$-count
\cite{Amy2014Polynomial-Time,Beverland2019Lower-bounds-on,Heyfron_2018,kissinger2019reducing},
a metric of some importance when considering fault-tolerant quantum
computation.  However, since we consider raw physical circuits, the
metrics of interest for us are the total circuit depth and the number
of two-qubit gates, since minimising these parameters serves as a good
proxy for minimising total error rate in the circuit.  The novel
contribution is a new technique for circuit optimisation by exploiting
symmetric structures for exponentials of Pauli strings, called
\emph{Pauli gadgets}, derived using phase gadget structures in the
\zxcalculus.  Pauli gadgets occur naturally in quantum simulations
where a Hamiltonian is decomposed into a sum of Pauli tensors and
Trotterised \cite{1367-2630-18-2-023023}.  Hence these techniques are
specifically useful to optimise quantum circuits for quantum chemistry
simulations \cite{Cao:2018aa}.

\begin{figure}
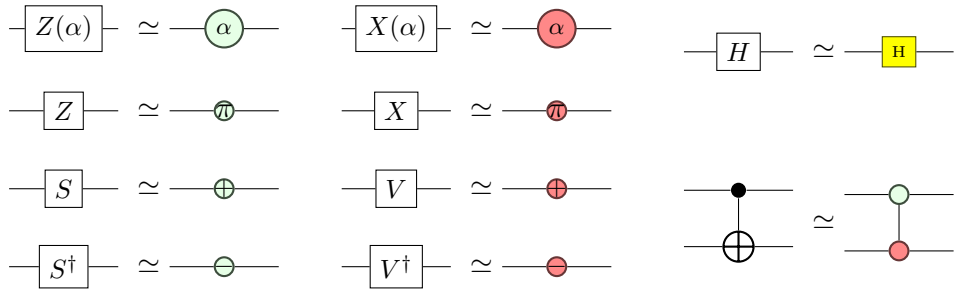

\centering
\begin{tabular}{ccc}
\begin{minipage}{.25\textwidth}
\begin{equation*}
\inltf{GatesRZ} \simeq \inltf{ZXCalcRZ}
\end{equation*}
\end{minipage} &
\begin{minipage}{.25\textwidth}
\begin{equation*}
\inltf{GatesRX} \simeq \inltf{ZXCalcRX}
\end{equation*}
\end{minipage} &
\multirow{2}{*}{
\begin{minipage}{.25\textwidth}
\begin{equation*}
\inltf{GatesH} \simeq \inltf{ZXCalcH}
\end{equation*}
\end{minipage}
}
\\
\begin{minipage}{.25\textwidth}
\begin{equation*}
\inltf{GatesZ} \simeq \inltf{ZXCalcZ}
\end{equation*}
\end{minipage} &
\begin{minipage}{.25\textwidth}
\begin{equation*}
\inltf{GatesX} \simeq \inltf{ZXCalcX}
\end{equation*}
\end{minipage} & \\
\begin{minipage}{.25\textwidth}
\begin{equation*}
\inltf{GatesS} \simeq \inltf{ZXCalcS}
\end{equation*}
\end{minipage} &
\begin{minipage}{.25\textwidth}
\begin{equation*}
\inltf{GatesV} \simeq \inltf{ZXCalcV}
\end{equation*}
\end{minipage} &
\multirow{2}{*}{
\begin{minipage}{.25\textwidth}
\begin{equation*}
\inltf{GatesCX} \simeq \inltf{ZXCalcCX}
\end{equation*}
\end{minipage}
}
\\
\begin{minipage}{.25\textwidth}
\begin{equation*}
\inltf{GatesSdg} \simeq \inltf{ZXCalcSdg}
\end{equation*}
\end{minipage} &
\begin{minipage}{.25\textwidth}
\begin{equation*}
\inltf{GatesVdg} \simeq \inltf{ZXCalcVdg}
\end{equation*}
\end{minipage} &
\end{tabular}
\caption{Common circuit gates and their representations in the scalar-free \zxcalculus}
\label{fig:zx_gates}
\end{figure}

\noindent
\textbf{Notation:\ } In the following, we will mix freely the usual quantum circuit
notation and the scalar-free \zxcalculus \cite{Coecke:2009aa}. For both 
forms of diagram, we will follow a left-to-right convention. We will also adopt 
the same convention for composition of circuits in equations, i.e. $C ; D$ 
means we apply $C$ first, followed by $D$. A
translation of common gates between the two formalisms is given in
Figure \ref{fig:zx_gates}.  A brief introduction to the \zxcalculus
is found in \cite{fagan2019optimising}; for a complete treatment see
\cite{Coecke2017Picturing-Quant}.  For reasons of space we omit the
\zxcalculus inference rules, however we use the complete set of
Vilmart \cite{vilmart2018near}. 

\noindent
\textbf{Remark:\ } Late during the preparation of this paper, it came
to our attention that Litinski~\cite{litinski2018game}
has defined a notation for Pauli product operators essentially
equivalent to the Pauli gadgets of Section~\ref{sec:exponentials}.
Since that work concerns computing under a surface code, this suggests
applications of our approach beyond the near term quantum devices we
focus on here.  The use of \zxcalculus for lattice surgery by de
Beaudrap and Horsman \cite{Beaudrap2017The-ZX-calculus} offers an
obvious route.

\section{Circuit Optimisations}
\label{sec:gen_optimisation}

Circuit optimisation is typically carried out by pattern replacement:
recognising a subcircuit of specific form and replacing it with an
equivalent.  This is sometimes called \emph{peephole optimisation} in
analogy to local optimisation techniques in classical compilers;
however in the case of quantum circuits any connected subcircuit can
be replaced, including the entire circuit.  Usually the replacement is
cheaper with respect to some cost metric, but in a multi-pass
optimiser like \tket, the replacement may enable a more powerful later
optimisation pass, rather than improving the circuit itself, or map
the circuit onto a particular gate set supported by the target device.

In \tket, circuits are represented internally as non-planar maps, a
generalisation of directed graphs wherein the incident edges at each
vertex are ordered, to admit non-commutative operations like the \CNot
gate.  Unlike operation lists or discrete time frames, this
representation preserves only the connectivity of the operations,
abstracting away qubit permutations and timing information.  The \tket
optimiser consists of multiple rewriting strategies called
\emph{passes} which may be combined to achieve the desired circuit
transformation\footnote{
  We regret that at the time of writing this feature is not in the
  publicly available \texttt{pytket} release; it is planned for a
  future release.
}.  The primitive rewriting steps are computed by the
double pushout method \cite{Ehrig:2006ab}, although the matching is
usually achieved by a custom search algorithm for efficiency reasons.

Simple examples include merging adjacent rotation gates acting on the
same basis, cancelling operation-inverse pairs, and applying
commutation rules.  Any sequence of single-qubit operations may be
fused into a single unitary, for which an Euler decomposition can be
computed.  \tket has the possibility to choose which basis of
rotations to use for the Euler form -- for example $ZXZ$ or $XZX$ --
depending on local context, which can permit more commutations, or
easy translation to a native gate set (for example, $ZYZ$
triples are useful to match the U3 gate supported in the Qiskit
framework \cite{qiskit}).

If the circuit contains a long sequence of gates acting on the same
two qubits, the $KAK$ (Cartan) decomposition
\cite{blaauboer2008analytical,vidal2004universal} may be applied.
This gives a canonical form requiring at most three \CNot gates.
Even when arbitrary rotations are permitted, realistic circuits
include significant Clifford subcircuits.  In particular, \tket takes
rules from \cite{fagan2019optimising} to reduce any pair of \CNot
gates that are separated only by single-qubit Clifford gates.  However
there is a very wide literature on Clifford circuits which could be
applied here
\cite{amy2016finite,Duncan:2019aa,selinger2013generators}. 
In the following sections we describe a novel technique for optimising
a new class of multi-qubit subcircuits, called phase gadgets and Pauli gadgets.

\section{Phase Gadgets}
\label{sec:phase_gadgets}

In principle, local rewriting of gate sequences is sufficient for any
circuit optimisation\footnote{This is a consequence of the
  completeness of the \zxcalculus \cite{vilmart2018near}.}. However, in practice, good results often require
manipulation of large-scale structures in the quantum circuit.
\emph{Phase gadgets} are one such macroscopic structure that is easy to
identify within circuits, easy to synthesise back into a circuit, and
have a useful algebra of interactions with one another.

\begin{definition}\label{def:phasegadget}
  The \emph{$Z$-phase gadgets} $\Phi_n(\alpha) : \mathbb{C}^{\otimes n}
  \to \mathbb{C}^{\otimes n}$ are a family of unitary maps we define
  recursively as :
  \[
  \Phi_1(\alpha) :=  Z(\alpha)
  \qquad\qquad
  \Phi_{n+1}(\alpha) :=
  (\CX \otimes 1_{n-1}) ;
  (1_1 \otimes \Phi_n(\alpha)) ;
  (\CX \otimes 1_{n-1})
  \]
  \[
  \inltf{GadgetDef-lhs} 
  \quad = \quad 
  \inltf{GadgetDef-rhs} 
  \]
\end{definition}

\begin{remark}
  We could equally define the $X$-phase gadget as the colour dual of
  the $Z$-phase gadget, and the $Y$-phase gadget by conjugating the
  $Z$-phase gadget with $X(\frac{\pi}{2})$ rotations.  Since we won't
  needs these in this paper, we'll refer to the $Z$-phase gadget
  simply as a phase gadget.
\end{remark}

\begin{lemma}\label{lem:phasegadgetzx}
  In \zxcalculus notation we have:
  \[
  \Phi_n(\alpha) := \inltf{PhaseGadgetDef}
  \]
\end{lemma}

\begin{corollary}\label{cor:basic-phsgad-eqns}
  We have the following laws for decomposition, commutation, and
  fusion of phase gadgets.
  \begin{align*}
{\inltf{PhaseGadgetCNOT-lhs}} & \quad = \quad
{\inltf{PhaseGadgetCNOT}}   \\ \\
{\input{PhaseGadgetCommute0.tikz}} & \quad = \quad
{\input{PhaseGadgetCommute1.tikz}} \\ \\
{\input{PhaseGadgetFusion0.tikz}} & \quad = \quad
{\input{PhaseGadgetFusion1.tikz}}
  \end{align*}
\end{corollary}

\begin{figure}[ht]
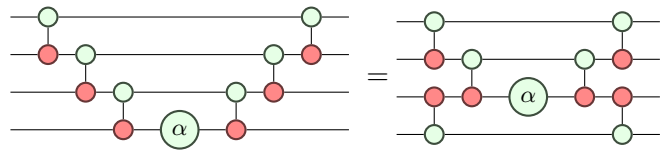

\begin{equation*}
  {\inltf{PhaseGadgetLadder}} = {\inltf{PhaseGadgetTree}}
\end{equation*}
\caption{Comparing the worst-case and best-case patterns for constructing phase gadgets with respect to \CNot depth. The left shows a \CNot ladder as produced within the Unitary Coupled Cluster generator in IBM Qiskit Aqua, and the right is the optimal balanced-tree form used by \tket.}
\label{fig:ladder_vs_tree}
\end{figure}

The decomposition law gives the canonical way to synthesise a quantum
circuit corresponding to a given phase gadget.  However, from the
\zxcalculus form, it's immediate that phase gadgets are invariant
under permutation of their qubits, giving the compiler a lot of
freedom to synthesise circuits which are amenable to optimisation.
As a simple example, the naive \CNot ladder approach, shown in
Figure~\ref{fig:ladder_vs_tree}, requires a \CNot-depth of $2(n-1)$ to
synthesise an $n$-qubit phase gadget; replacing this with a balanced
tree yields a \CNot depth of $2\lceil \log n \rceil$. Note that the
quantity of \CNot gates used is still (and always will be) $2(n-1)$, but we
can still obtain benefits with respect to depth.

Further, in the balanced tree form more of the \CNot gates are
``exposed'' to the rest of the circuit, and could potentially be
eliminated by a later optimisation pass.  Note that this form is not
unique, allowing synthesis informed by the circuit context in which the
phase gadget occurs.  For example, \tket aligns the \CNot{}s between
consecutive phase gadgets whenever possible.

Trotterised evolution operators, as commonly found in quantum
chemistry simulations, have the general form of a sequence of phase
gadgets, separated by a layer of single-qubit Clifford rotations.
For each consecutive pair of gadgets, if the outermost \CNot{}s align
then they can both be eliminated, or if there are some intervening
Clifford gates then we can use Clifford optimisation techniques to
remove at least one of the \CNot{}s.

\section{Pauli Gadgets}
\label{sec:exponentials}

In the language of matrix exponentials, the phase gadget
$\Phi_n(2\alpha)$  corresponds to the operator $e^{-i\alpha
  Z^{\otimes n}}$. 
A consequence of Corollary~\ref{cor:basic-phsgad-eqns} is that any
circuit $P$ consisting entirely of $Z$-phase gadgets can be represented
succinctly in the form:
\begin{equation}
  P\ket{x_1 x_2 \ldots x_n} = e^{-i \sum_j \alpha_j f_j(x_1, x_2, \ldots, x_n)} \ket{x_1 x_2 \ldots x_n}
\end{equation}
for some Boolean linear functions $f_j$.  For comparison,
\emph{phase-polynomial} circuits $C$ (the class of circuits that can be built from $\{\CNot, T\}$ \cite{amy2013meet}) can be represented as:
\begin{equation}
  C\ket{x_1 x_2 \ldots x_n} = e^{i\frac{\pi}{4}\sum_j f_j(x_1, x_2, \ldots, x_n)} \ket{g(x_1, x_2, \ldots, x_n)}
\end{equation}
for Boolean linear functions $f_j$ and a linear reversible function
$g$. There is already a wide literature covering phase-polynomials
and optimisations with them \cite{Amy2014Polynomial-Time,amy2019t,Nam:2018aa}.

\begin{figure}[th]
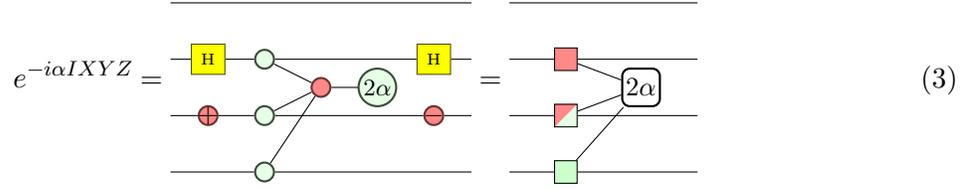

\begin{equation}\label{eq:pauli-exp-def}
e^{-i\alpha I X Y Z} = {\input{PhaseGadgetIXYZ.tikz}} = {\input{PauliExpDef.tikz}}
\end{equation}
\caption{An example of the correspondence between Pauli evolution operators and phase gadgets. We introduce this notation on the right for a more succinct graphical representation. The green, red, and mixed-colour boxes respectively represent the Pauli gadget acting on a qubit in the $Z$, $X$, and $Y$ bases. These are formed by a phase gadget on the qubits (generating all $Z$ interactions), then optionally conjugating the qubits with Hadamard gates for $X$, or $X(\frac{\pi}{2})$ gates for $Y$. We omit trivial qubits ($I$) from the diagrammatic representation.}
\label{fig:pauli_exp_gadget}
\end{figure}

The correspondence between phase gadgets and matrix exponentials
generalises to exponentials of any Pauli tensor $e^{-i\alpha \sigma_1
  \sigma_2 \ldots \sigma_n}$, by conjugating the phase gadget with
approriate Clifford operators as shown in
Figure~\ref{fig:pauli_exp_gadget}.  

\begin{definition}\label{def:pauli-exp}
  Let $s$ be a word over the alphabet
  $\{X,Y,Z\}$; then the \emph{Pauli gadget}
  $P(\alpha,s)$ is defined as $U(s) ; \Phi_{\sizeof{s}}(\alpha) ;
  U(s)^\dag$ where the unitary $U(s)$ is defined by recursion over
  $s$:
  \[
  U(Z s') = I \otimes U(s')   \qquad
  U(Y s') = X(\frac{\pi}{2})  \otimes U(s') \qquad
  U(X s') = H  \otimes U(s')
  \]
\end{definition}
\TODO{Do we need a base case for the empty string here?}

\noindent
Definition~\ref{def:pauli-exp}  can be easily extended to aribitrary strings over the
Paulis (\ie including the identity) by adding wires which the phase
gadget does not act on.  This is illustrated in
Figure~\ref{fig:pauli_exp_gadget}.  Taking advantage of this we'll
generally assume that the Pauli gadget is the full width of the
circuit, although it may not act on every qubit.

In general, Pauli gadgets present difficulties for
phase-polynomial-based circuit optimisation methods, as not all pairs
of Pauli evolution operators will commute (for the simplest example,
consider $e^{-i\alpha X} e^{-i\beta Z} \neq e^{-i\gamma
  Z} e^{-i\delta X}$ for any non-degenerate choices of
angles).  We now generalise the results of the preceding section to
consider interactions between Pauli gadgets.  The following is
easy to demonstrate using matrix exponentials.

\begin{proposition}\label{prop:gen-euler}
  Let $P$ and $Q$ be Pauli tensors, then either (i)  $e^{-i\alpha
    P}e^{-i\beta Q} = e^{-i\beta Q} e^{-i\alpha P}$ for all $\alpha$
  and $\beta$;  or (ii)  for all $\alpha_i$ there exist $\beta_i$ such
  that
  \begin{equation}\label{eq:gen-euler}
    e^{-i\alpha_1 P} e^{-i\alpha_2 Q} e^{-i\alpha_3 P}
    = e^{-i\beta_1 Q} e^{-i\beta_2 P} e^{-i\beta_3 Q}
  \end{equation}
\end{proposition}

\noindent
Note that the $\alpha_i$ and $\beta_i$ are computed as the Euler-angle
decompositions of a combined rotation.  Taking $P = Z$ and $Q =
X$, Equation~\eqref{eq:gen-euler} is axiom (EU) of the
ZX-calculus \cite{vilmart2018near}.  We will give a
\zxcalculus proof of this theorem for Pauli gadgets, with an
intermediate state giving a very compact circuit representation for
any consecutive pair of Pauli gadgets.

The following lemmas have elementary proofs.

\begin{lemma}\label{lem:1qb-cliff-rules}
  The commutation rules for Pauli gadgets and single-qubit
  Clifford gates, shown in
  Figure~\ref{fig:exponential_clifford_rules} are derivable in the
  \zxcalculus.
\end{lemma}
\TODO{Proof in appendix}

\begin{figure}[t!]
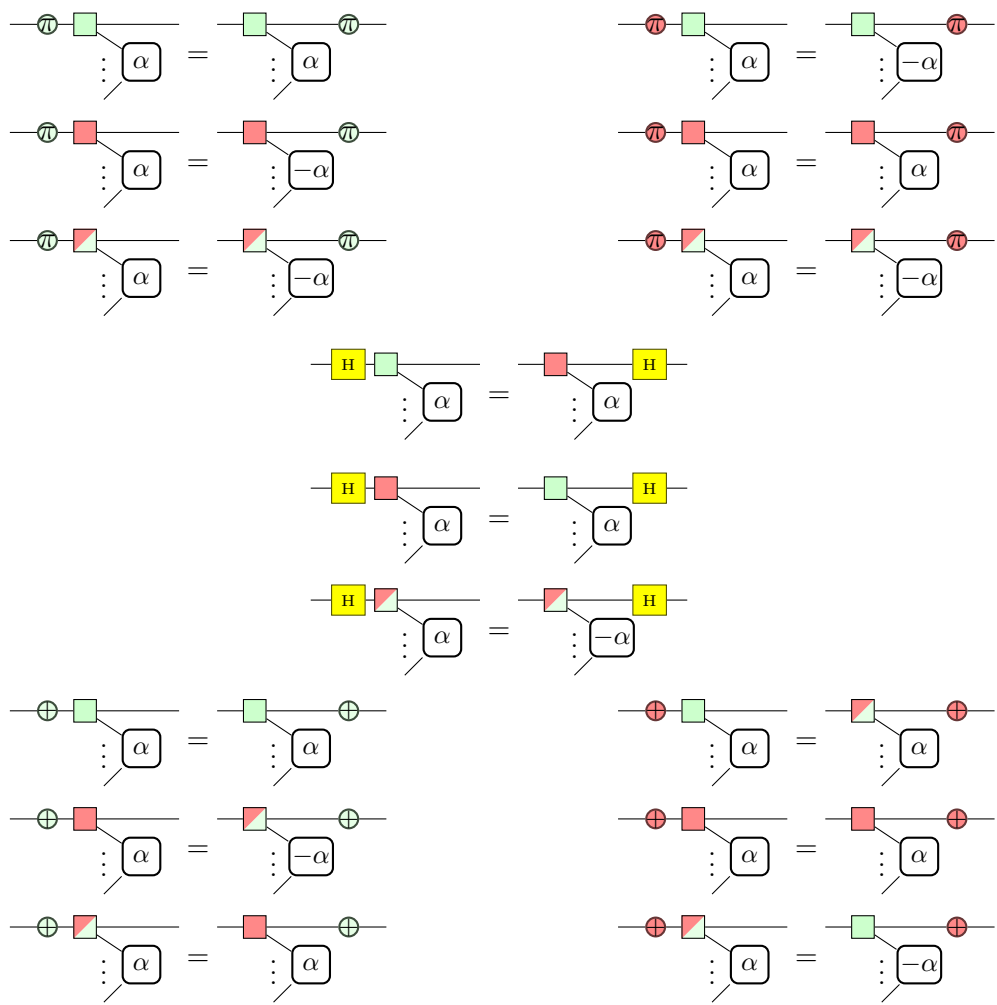

\begin{subfigure}[b]{0.5\textwidth}
\begin{equation*}
{\input{PauliExpZZa.tikz}} = {\input{PauliExpZZb.tikz}}
\end{equation*}
\begin{equation*}
{\input{PauliExpZXa.tikz}} = {\input{PauliExpZXb.tikz}}
\end{equation*}
\begin{equation*}
{\input{PauliExpZYa.tikz}} = {\input{PauliExpZYb.tikz}}
\end{equation*}
\end{subfigure}
\begin{subfigure}[b]{0.5\textwidth}
\begin{equation*}
{\input{PauliExpXZa.tikz}} = {\input{PauliExpXZb.tikz}}
\end{equation*}
\begin{equation*}
{\input{PauliExpXXa.tikz}} = {\input{PauliExpXXb.tikz}}
\end{equation*}
\begin{equation*}
{\input{PauliExpXYa.tikz}} = {\input{PauliExpXYb.tikz}}
\end{equation*}
\end{subfigure}
\begin{equation*}
{\input{PauliExpHZa.tikz}} = {\input{PauliExpHZb.tikz}}
\end{equation*}
\begin{equation*}
{\input{PauliExpHXa.tikz}} = {\input{PauliExpHXb.tikz}}
\end{equation*}
\begin{equation*}
{\input{PauliExpHYa.tikz}} = {\input{PauliExpHYb.tikz}}
\end{equation*}
\begin{subfigure}[b]{0.5\textwidth}
\begin{equation*}
{\input{PauliExpSZa.tikz}} = {\input{PauliExpSZb.tikz}}
\end{equation*}
\begin{equation*}
{\input{PauliExpSXa.tikz}} = {\input{PauliExpSXb.tikz}}
\end{equation*}
\begin{equation*}
{\input{PauliExpSYa.tikz}} = {\input{PauliExpSYb.tikz}}
\end{equation*}
\end{subfigure}
\begin{subfigure}[b]{0.5\textwidth}
\begin{equation*}
{\input{PauliExpVZa.tikz}} = {\input{PauliExpVZb.tikz}}
\end{equation*}
\begin{equation*}
{\input{PauliExpVXa.tikz}} = {\input{PauliExpVXb.tikz}}
\end{equation*}
\begin{equation*}
{\input{PauliExpVYa.tikz}} = {\input{PauliExpVYb.tikz}}
\end{equation*}
\end{subfigure}
\caption{Rules for passing Clifford gates through Pauli gadgets.}
\label{fig:exponential_clifford_rules}
\end{figure}

\begin{lemma}\label{lem:cnot-rules}
  The commutation rules for Pauli gadgets and \CX gates, shown in
  Figure~\ref{fig:exponential_cx_rules} are derivable in the
  \zxcalculus.
\end{lemma}
\TODO{Proof in appendix}

\begin{figure}[t!]
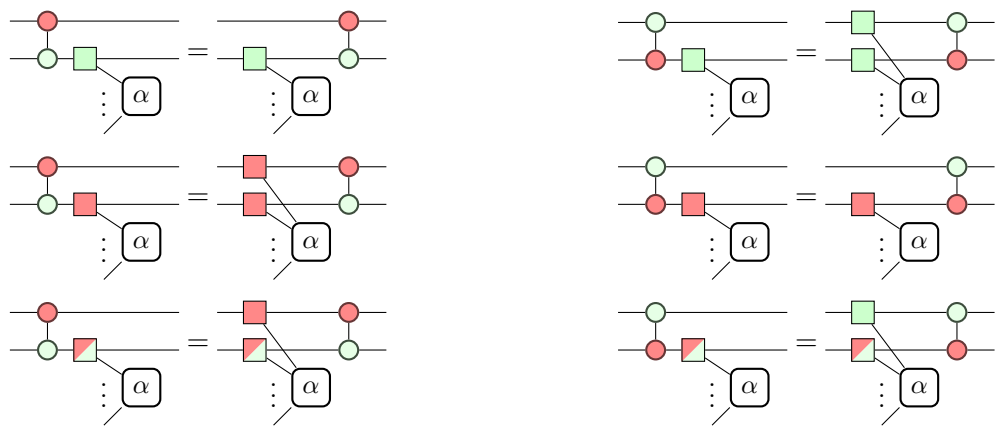

\begin{subfigure}[b]{0.5\textwidth}
\begin{equation*}
{\input{PauliExpCXZga.tikz}} = {\input{PauliExpCXZgb.tikz}}
\end{equation*}
\begin{equation*}
{\input{PauliExpCXXga.tikz}} = {\input{PauliExpCXXgb.tikz}}
\end{equation*}
\begin{equation*}
{\input{PauliExpCXYga.tikz}} = {\input{PauliExpCXYgb.tikz}}
\end{equation*}
\end{subfigure}
\begin{subfigure}[b]{0.5\textwidth}
\begin{equation*}
{\input{PauliExpCXZra.tikz}} = {\input{PauliExpCXZrb.tikz}}
\end{equation*}
\begin{equation*}
{\input{PauliExpCXXra.tikz}} = {\input{PauliExpCXXrb.tikz}}
\end{equation*}
\begin{equation*}
{\input{PauliExpCXYra.tikz}} = {\input{PauliExpCXYrb.tikz}}
\end{equation*}
\end{subfigure}
\caption{Rules for passing \CX gates through Pauli gadgets.}
\label{fig:exponential_cx_rules}
\end{figure}

\noindent
Note that Figures~\ref{fig:exponential_clifford_rules} and
\ref{fig:exponential_cx_rules} are not exhaustive, but they
suffice for present purposes.

It will be useful to define some notation for working with strings of
Paulis.  For strings $s$ and $t$ we write their concatenation as $st$;
$s_i$ denotes the $i$th symbol of $s$; and $\sizeof{s}$ denotes the
length of $s$.  A string consisting entirely of $I$ is called
\emph{trivial}.  We say that $t$ is a \emph{substring} of $s$ when,
for all $i$, $s_i \neq t_i$ implies $t_i = I$; if in addition $s \neq
t$ and $t$ is non-trivial then $t$ is \emph{proper substring}.  We
write $t\bullet s$ for the pointwise multiplication of Pauli strings
(up to global phase); in particular if $t$ is a substring of $s$ then
$(t\bullet s)_i = I$ iff $s_i = t_i$ and is $s_i$ otherwise.  The
\emph{intersection} of strings $s$ and $t$ is the set of indices $i$
satisfying $s_i \neq I$ and $t_i \neq I$.

\begin{lemma}\label{lem:reduce-paulexps}
  Let $st$ be a Pauli string; then for all $\alpha$ there exists a
  Clifford unitary $U$ acting on $\sizeof{s} + 1$ qubits such that
  \[
  P(\alpha,st)  = (U \otimes I_{\sizeof{t -1}})
  ; (I_{\sizeof{s}}\otimes P(\alpha,t))
  ; (U^\dag \otimes I_{\sizeof{t -1}})
  \]
  \[
  \inltf{PauliSubstring-lhs} 
  \qquad = \qquad
  \inltf{PauliSubstring-rhs} 
  \]
  Further, $U$ can be constructed in a canonical form which depends
  only on the string $s$.
  \TODO{It is still dependent on one symbol of $t$ - if the chosen qubit of $P(\alpha,t)$ is $X$, then the \CX ladder will commute through and eliminate}
  \begin{proof}
    For simplicity of exposition we assume $s_i, t_j \neq I$ for all
    $i$ and $j$.  We construct $U$ in two layers. The first layer of gates corresponds 
    to $U(s)$ from Definition~\ref{def:pauli-exp}. By
    \ref{lem:1qb-cliff-rules}, these gates can pass through $P(\alpha,Z^{|s|}t)$ and 
    cancel with their inverses from $U^\dagger$ to give $P(\alpha,st)$. 
    Similarly, a \CNot gate on the first two qubits can pass through $I \otimes 
    P(\alpha,Z^{|s|-1}t)$ to give $P(\alpha,st)$ by Lemma~\ref{lem:cnot-rules}. 
    The second layer of $U$ is a chain of \CNot gates that repeats this to 
    convert $P(\alpha,t)$ to $P(\alpha,Z^{|s|}t)$. The final \CNot in this 
    chain acts has its target on the $(|s|+1)$th qubit, corresponding to $t_1$. 
    If $t_1=X$, then the \CNot will commute through $P(\alpha,t)$ without 
    extending it, so additional single qubit gates may be required around the 
    \CNot to map $t_1$ to $Z$ and back. Composing these layers gives a $U$ that 
    can pass through $I_{|s|} \otimes P(\alpha,t)$ and cancel with $U^\dagger$ 
    to leave $P(\alpha,st)$.
  \end{proof}
\end{lemma}

\begin{remark}\label{rem:choice-of-cnots}
  As shown in \ref{fig:ladder_vs_tree}, the \CNot part of $U$ may be
  more efficiently constructed as a balanced tree, or some other
  configuration which allows later gate cancellation.
\end{remark}

\begin{corollary}\label{cor:reduce-pauli-gads}
  Let $t$ be a proper substring of $s$; then there exists a unitary
  $U$ and a permutation $\pi$ such that
  \[
  P(\alpha,s)  =
  \pi ; (U \otimes I_{|s|-|t|-1}) ; \pi^\dag  ;
  P(\alpha,t\bullet s) ;
  \pi^\dag ; (U^\dag \otimes I_{|s|-|t|-1}) ; \pi
  \]
\end{corollary}

\begin{corollary}\label{cor:same-paulis-fuse}
  Let $s$ be a Pauli string; then for all $\alpha$ and $\beta$:
  \[
  P(\alpha,s) ; P(\beta,s) = P(\alpha + \beta,s)
  \]
\end{corollary}

\begin{lemma}\label{lem:intermediate-form}
  Let $s$ and $t$ be Pauli strings; then there exists a Clifford
  unitary $U$ such that
  \[
  P(\alpha,s); P(\beta,t) = U ; P(\alpha,s') ; P(\beta,t')
  ; U^\dag
  \]
  where $s'$ and $t'$ are Pauli strings with intersection at most 1.
  \begin{proof}
    Let $r$ denote the maximum common substring of $s$ and $t$.  Then
    by Corollary~\ref{cor:reduce-pauli-gads} we have
    \begin{equation}\label{eq:eliminate_common}
    P(\alpha,s); P(\beta,t)
    = 
    U_r ; P(\alpha,r\bullet s) ;
    U_r^\dag ; U_r ; P(\beta,r\bullet t) ; U_r^\dag
    =
    U_r ; P(\alpha,r\bullet s) ; P(\beta,r\bullet t) ; U_r^\dag
    \end{equation}
    hence we will assume that $s$ and $t$ have no non-trivial common
    substring.  Now suppose that $s_i = Y$ and $t_i = X$.  Applying
    Lemma~\ref{lem:1qb-cliff-rules} we can replace $s_i$ with a $Z$
    node by conjugating with $X(\pi/2)$; since $X$ rotations commute
    with $X$ nodes, this unitary can move outside the two gadgets.
    \begin{equation}
    \inltf{PauliExp-YX-replace-i} 
    \quad = \quad
    \inltf{PauliExp-YX-replace-ii} 
    \quad = \quad
    \inltf{PauliExp-YX-replace-iii} 
    \end{equation}
    The pairing of $Y$ and $Z$ can be treated the same way.  Hence we
    can assume that the symbol $Y$ does not occur in the intersection
    of $s$ and $t$.

    Now we proceed by induction on the size of the intersection.  If
    the intersection is size 0 or 1 then we have the result.
    Otherwise consider two non-trivial qubits $i$ and $j$ in the
    intersection.  Suppose $s_i = s_j = Z$ and $t_i = t_j = X$; then
    by Lemma \ref{lem:cnot-rules} we can reduce the size of the intersection by
    two as shown below:
    \begin{equation}\label{eq:eliminate_mismatch}
    \inltf{XXZZ-reduce-i}
    \quad = \quad
    \inltf{XXZZ-reduce-ii}
    \quad = \quad
    \inltf{XXZZ-reduce-iii}
    \end{equation}
    The only other case to be considered is when $s_i = t_j = X$ and $s_j
    = t_i = Z$, in which case Lemma \ref{lem:1qb-cliff-rules} gives the following
    reduction.
    \begin{equation}
    \inltf{XZXZ-reduce-i}
    \quad = \quad
    \inltf{XZXZ-reduce-iib}
    \quad = \quad
    \inltf{XZXZ-reduce-iii}
    \end{equation}
    Hence the size of the intersection can be reduced to less than two.
  \end{proof}
\end{lemma}

\begin{theorem}\label{thm:pauli-exp-euler}
  Let $s$ and $t$ be strings of Paulis. Either the corresponding
  gadgets commute:
  \[
  \forall \alpha, \beta \quad P(\alpha,s) ; P(\beta,t) = P(\beta,t) ; P(\alpha,s)
  \]
  or they satisfy the Euler equation:
  \[
  \forall \alpha_i, \exists \beta_i  \quad
  P(\alpha_1,s) ; P(\alpha_2,t) ; P(\alpha_3,s)
  = P(\beta_1,t) ; P(\beta_2,s) ; P(\beta_3,t)
  \]
  \begin{proof}
    By Lemma~\ref{lem:intermediate-form}, we have $U$, $s'$ and $t'$
    such that  
    \[
    P(\alpha,s) ; P(\beta,t) = 
    U ; P(\alpha,s') ; P(\beta,t') ; U^\dag
    \]
    Where $s'$ and $t'$ have at most intersection 1.  If their
    intersection is trivial, or if both gadgets act on their common
    qubit in the same basis (Corollary~\ref{cor:same-paulis-fuse}),  
    then they commute, from which we have
    \begin{equation}
    U ; P(\alpha,s') ; P(\beta,t') ; U^\dag
    =
    U ; P(\beta,t') ; P(\alpha,s') ; U^\dag
    =
    P(\beta,t) ; P(\alpha,s)
    \end{equation}
    Otherwise the gadgets need not commute, but the Euler equation
    holds.  Without loss of generality assume that $s$ is all $Z$s and
    $t$ is all $X$s. In the case where $|s| = |t| = 2$, we continue as follows:
    
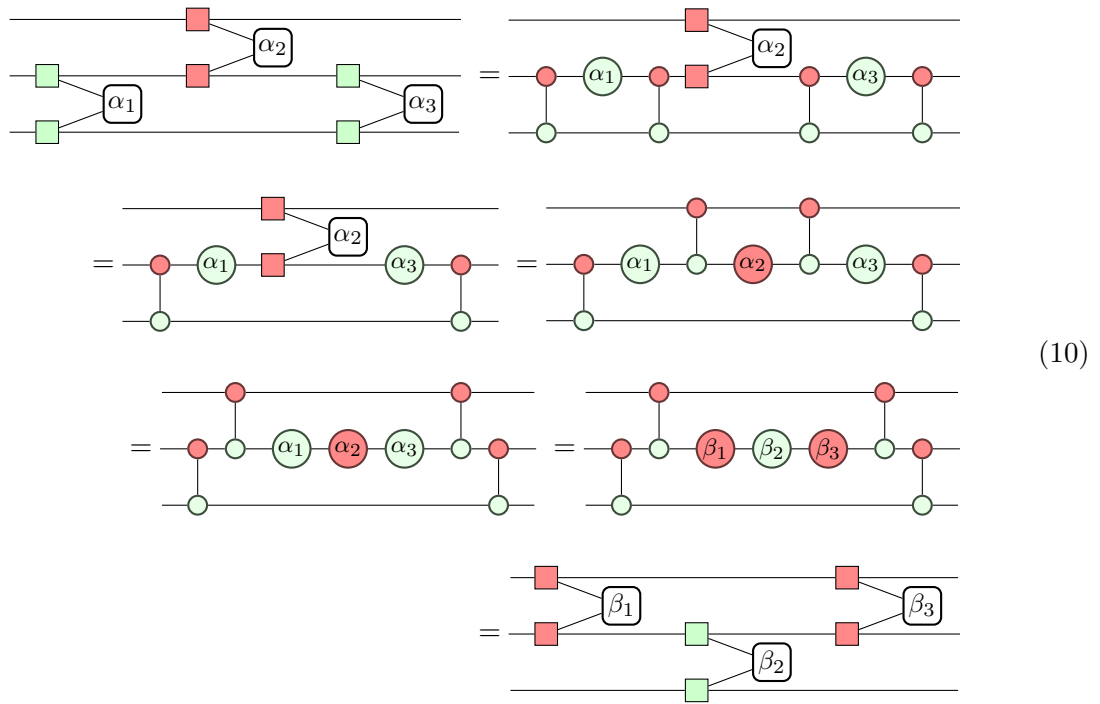
\begin{equation}
    \begin{aligned}
    {\inltf{PauliExpSingleEul2a}} 
    = {\inltf{PauliExpSingleEul2b}} \\\\ 
    = {\inltf{PauliExpSingleEul2bi}}
    = {\inltf{PauliExpSingleEul2ci}} \\\\ 
    = {\inltf{PauliExpSingleEul2c}}
    = {\inltf{PauliExpSingleEul2d}} \\\\
    = {\inltf{PauliExpSingleEul2e}}
    \end{aligned}
    \end{equation}
    This applies Lemma~\ref{lem:cnot-rules} to decompose Pauli gadgets and 
    commute \CNot gates, followed by the (EU) rule and essentially reversing 
    the procedure. This generalises to larger $s$ and $t$ by applying Lemma~\ref{lem:reduce-paulexps}.
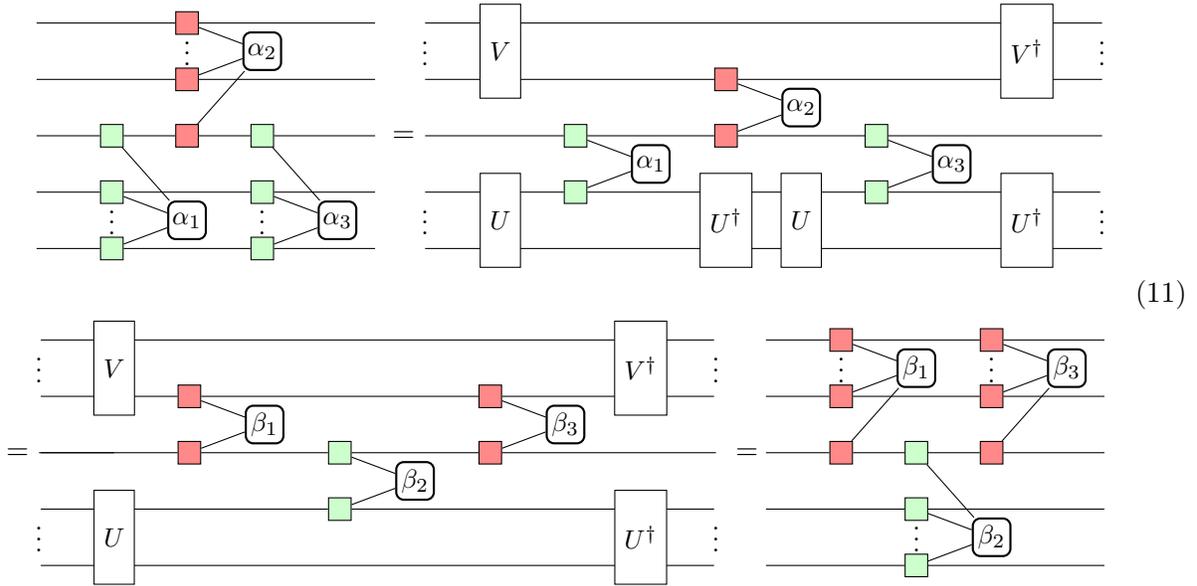
    \begin{equation}
    \begin{aligned}
      {\inltf{PauliExpSingleEula}} 
      = {\inltf{PauliExpSingleEulb}} \\\\
      = {\inltf{PauliExpSingleEulf}} 
      = {\inltf{PauliExpSingleEulg}}
    \end{aligned}
  \end{equation}
  \end{proof}
\end{theorem}

Synthesising a Pauli gadget $P(\alpha,s)$ in isolation requires $2(|s|-1)$ \CNot gates, hence $P(\alpha,s) ; P(\beta,t)$ would usually require $2(|s|+|t|-2)$ \CNot gates in total. Applying Equation~\ref{eq:eliminate_common} will reduce the total cost by $2$ for each qubit in the maximum common substring. Equation~\ref{eq:eliminate_mismatch} uses two gates to reduce the gadgets by 1 qubit each, giving a net saving of $2$ \CNot gates per application. This reduces the total cost to $2(|s| + |t| - |r| - \lfloor \frac{|u|}{2} \rfloor - 2)$ \CNot gates where $r$ is the maximum common substring of $s$ and $t$, and $u$ is the subset of the intersection of $s$ and $t$ that is not in $r$. In the case where $s$ and $t$ act on the same set of qubits and $|s\bullet t| \leq 2$, we can synthesise the pair $P(\alpha,s) ; P(\beta,t)$ using the same number of \CNot{}s as just $P(\alpha,s)$. Performance with respect to depth is harder to assess analytically and will be left for future work.


\section{Optimisation Example}
\label{sec:example}

The following example is a small region of a Unitary Coupled Cluster ansatz for 
analysing the ground state energy of $\text{H}_2$.   The parameters
$\alpha$ and $\beta$ are optimised by some variational method.
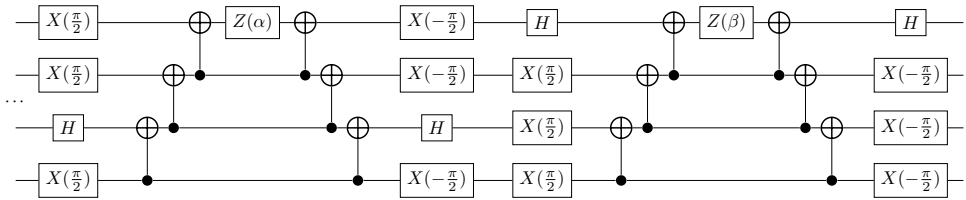
\begin{equation*}
\scalebox{0.7}{\input{H2Initial.tikz}}
\end{equation*}
The \CNot ladders in this circuit correspond to phase gadgets, so we 
start by detecting these and resynthesising them optimally to reduce the depth 
and expose more of the \CNot{}s to the rest of the circuit.
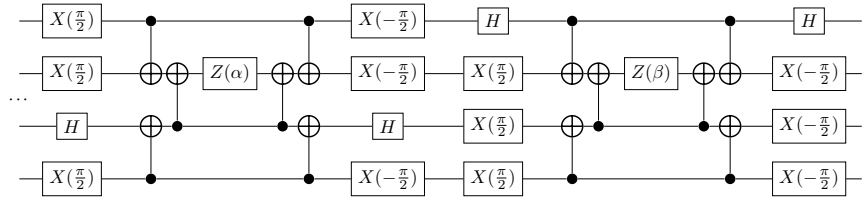
\begin{equation*}
\scalebox{0.7}{\input{H2PhaseSynth.tikz}}
\end{equation*}
Between the parametrised gates, there is a Clifford subcircuit, 
featuring some aligned \CNot pairs. The commutation 
and Clifford optimisation rules can further reduce the number of \CNot{}s here.
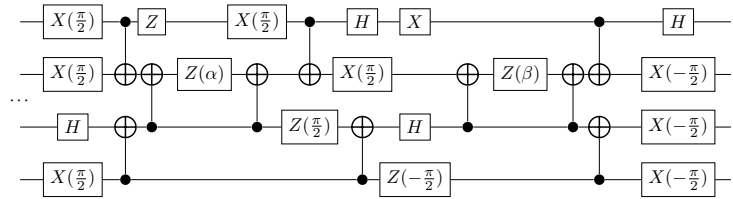
\begin{equation*}
\scalebox{0.7}{\input{H2Clifford2.tikz}}
\end{equation*}

Between phase gadget resynthesis and Clifford optimisations, we have 
successfully reduced the two-qubit gate count of this circuit from 12 to 10, 
and the depth with respect to two-qubit gates from 12 to 7. However, we could 
have noted that the original circuit corresponds to the operation $P(\alpha,YYXY) ; 
P(\beta,XYYY)$. These Pauli gadgets commute according to Theorem 
\ref{thm:pauli-exp-euler}. Following the proof, we can reduce the Pauli gadgets 
by stripping away the common qubits (where they both act on the $X$-basis) as in Equation~\ref{eq:eliminate_common}, and 
then reducing the remaining pair to simple rotations on different qubits using Equation~\ref{eq:eliminate_mismatch}. This 
yields an equivalent circuit using 6 two-qubit gates which can be arranged into 
only 4 layers.
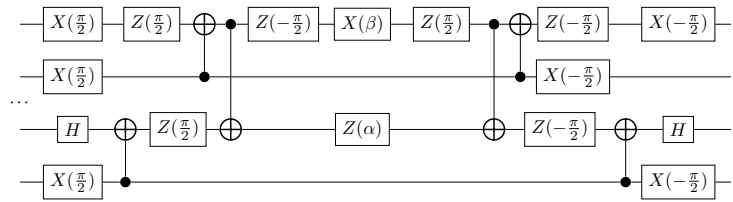
\begin{equation*}
\scalebox{0.7}{\input{H2Hand.tikz}}
\end{equation*}

\section{Results}
\label{sec:results}

Here we present some empirical results on the performance of these
optimisation techniques on realistic quantum circuits. We compared the
effectiveness of a few optimising compilers at reducing the number of
two-qubit interactions (\CNot or equivalent) in a circuit. For \tket,
we identified Pauli gadgets within the circuit and applied the aforementioned method for efficient pairwise synthesis, followed by Clifford subcircuit optimisation.

The test set used here consists of a small selection of circuits for
Quantum Computational Chemistry. They correspond to variational
circuits for estimating the ground state of small molecules
($\text{H}_{\text{2}}$, $\text{LiH}$, 
$\text{CH}_{\text{2}}$, or $\text{C}_{\text{2}}\text{H}_\text{4}$)
by the Unitary Coupled Cluster approach
\cite{barkoutsos2018quantum,bartlett1989alternative} using some
choice of qubit mapping (Jordan-Wigner \cite{jordan1928pauli}, Parity
mapping \cite{bravyi2017tapering}, or
Bravyi-Kitaev\cite{bravyi2002fermionic}) and chemical basis function
(sto-3g, 6-31g, cc-pvDZ, or cc-pvTZ). The bulk of each circuit is
generated by Trotterising some exponentiated operator, meaning many
phase/Pauli gadgets will naturally occur. These circuits were all
generated using the Qiskit Chemistry package \cite{qiskit} and the
QASM files can be found online \footnote{QASM files and the generating python script are available at: \url{https://github.com/CQCL/pytket/tree/master/examples/benchmarking/ChemistrySet}}.

\begin{table}
\centering
\begin{tabular}{lrr||rr|rr||rr} 
\multicolumn{3}{c||}{} & \multicolumn{2}{c|}{Qiskit 0.10.1} & \multicolumn{2}{c||}{PyZX} & \multicolumn{2}{c}{\makecell{CQC's \tket \\ 0.2}} \\
Name  & $g_{\rm in}$ & $d_{\rm in}$ & $g_{\rm out}$ &$d_{\rm out}$ & $g_{\rm out}$ & $d_{\rm out}$ & $g_{\rm out}$ & $d_{\rm out}$ \\ \hline
\begin{filecontents*}{ChemBenchResults-full.csv}
Test,CX count,CX depth,Pyquil count,Pyquil depth,Qiskit count,Qiskit depth,PyZX count,PyZX depth,Pytket count,Pytket depth
H2\_P\_sto3g,38,36,34,32,34,32,39,37,24,18
H2\_BK\_sto3g,38,38,32,32,36,36,43,40,25,23
H2\_JW\_sto3g,56,52,39,37,52,48,42,37,24,22
H2\_JW\_631g,768,744,,,624,600,467,392,349,324
LiH\_JW\_sto3g,8064,7616,,,6176,5776,6872,5579,3525,3173
LiH\_P\_sto3g,7640,7603,6355,6318,6222,6185,6838,5543,3992,3460
LiH\_BK\_sto3g,8680,8637,,,7402,7395,6577,5491,4226,4002
H2\_JW\_ccpvdz,14616,14436,,,10404,10224,8527,6105,5608,5396
CH2\_JW\_sto3g,21072,19749,,,15600,14565,18380,14995,9074,8064
LiH\_JW\_631g,69144,65676,,,49776,46596,58647,43835,27221,24462
H2\_JW\_ccpvtz,341280,339768,,,224640,223128,276691,151027,115376,113980
LiH\_JW\_ccpvdz,407320,388620,,,282608,264996,371847,229491,145714,134079
C2H4\_JW\_sto3g,640768,596857,,,438784,408313,-,-,248945,216796
\end{filecontents*}
\csvreader[
respect underscore, 
late after line=\\,
late after last line=,
]{ChemBenchResults-full.csv}
{1=\Name, 2=\Gates, 3=\Depth, 4=\gQuil, 5=\dQuil, 6=\gQis, 7=\dQis, 8=\gPyZX, 9=\dPyZX, 10=\gTket, 11=\dTket}
{\Name  &  \Gates & \Depth & \gQis & \dQis & \gPyZX & \dPyZX &  \gTket & \dTket }
\end{tabular}
\caption{Comparison of two-qubit gate count and depth for Quantum Computational Chemistry circuits achieved by quantum compilers. Each circuit was generated using a Unitary Coupled Cluster ansatz for ground state estimation of small molecules. The names of circuits indicate the molecule ($\text{H}_{\text{2}}$, $\text{LiH}$, $\text{CH}_{\text{2}}$, or $\text{C}_{\text{2}}\text{H}_\text{4}$), the qubit mapping (\textbf{J}ordan-\textbf{W}igner, \textbf{P}arity mapping, or \textbf{B}ravyi-\textbf{K}itaev), and chemical basis function (sto-3g, 6-31g, cc-pvDZ, or cc-pvTZ). $g_{in}$/$d_{in}$ denotes the two-qubit gate count/depth for the original circuits, and $g_{out}$/$d_{out}$ are the corresponding quantities for the optimised circuits from each compiler. Values correspond to the Pauli gadget optimisation pass in \tket, PyZX's \texttt{full\textunderscore reduce} procedure, and compilation with optimisation level 1 on Qiskit (at the time of writing, higher levels were found to not preserve the semantics of the circuit). Systems were allowed up to 10 hours of compute time for each circuit with timeouts indicated by blank cells.}
\label{tab:results-chemistry}
\end{table}

All of the implementations suffered from runtime scaling issues, meaning 
results for some of the larger circuits were reasonably unobtainable. Overall, \tket 
gained an average reduction of $54.5\%$ in \CNot count of the circuits, outperforming the $21.3\%$ from Qiskit and 
$16.3\%$ from PyZX. We find similar savings with respect to two-qubit gate depth, where 
\tket has an average reduction of $57.7\%$ ($21.8\%$ for Qiskit, $30.8\%$ for 
PyZX). This percentage is likely to improve as we start to look at even larger 
examples as the phase gadget structures are reduced from linearly-scaling \CNot 
ladders to the logarithmically-scaling balanced trees. We anticipate that 
incorporating the reduced form for adjacent Pauli gadgets will further cut down 
the \CNot count, especially given that rotations in the Unitary Coupled Cluster 
ansatz come from annihilation and creation operators, each generating a pair of 
rotations with very similar Pauli strings.

These empirical results were to compare pure circuit optimisation only, 
so no architectural constraints were imposed. It is left for future work to 
analyse how these techniques affect the ease of routing the circuit to conform 
to a given qubit connectivity map. This is non-trivial for the more macroscopic 
changes such as identifying and resynthesising phase gadgets which can change 
the interaction graph from a simple line to a tree. Recent work using Steiner 
trees \cite{kissinger2019cnot,nash2019quantum} could be useful for 
synthesising individual phase gadgets in an architecturally-aware manner.

As the quality of physical devices continues to improve, we can look 
forward to a future of fault-tolerant quantum computing. There has already been 
work making use of the structures discussed here in the domain of Clifford + T 
circuits. Notably, phase gadgets have found use recently for reducing the 
T-count of circuits \cite{kissinger2019reducing}. Another recent paper \cite
{litinski2018game} presents ways to usefully synthesise Clifford + T circuits 
in the realm of lattice surgery which use representations of rotations that are 
equivalent Pauli gadgets.


\small
\bibliography{opt}


\clearpage
\normalsize

\input{appendix.tex}

\end{document}

%% file: preamble.tex
\usepackage[T1]{fontenc}
\usepackage{lmodern}
\usepackage{fixltx2e}
\usepackage{microtype}
\usepackage{xspace}
\usepackage{enumitem}

\makeatletter
\newcommand\etc{etc\@ifnextchar.{}{.\@}\xspace}
\newcommand\ie{i.e.\@\xspace}  

\makeatother

\usepackage{graphicx}

\newcommand{\inlinegraphic}[2]{
  \dimendef\grafheight=255\dimendef\grafvshift=254
  \grafheight=#1
  \grafvshift=-0.5\grafheight
  \advance\grafvshift by 0.5ex
  \raisebox{\grafvshift}{\includegraphics[height=\grafheight]{images/#2}\xspace}
}

\newcommand{\ninlinegraphic}[2][1.0]{
  \dimendef\grafheight=255\dimendef\grafvshift=254
  \setbox0 = \hbox{\scalebox{#1}{\includegraphics{images/#2}}}
  \grafheight=\the\ht0
  \grafvshift=-0.5\grafheight
  \advance\grafvshift by 0.5ex
  \raisebox{\grafvshift}{\includegraphics[height=\grafheight]{images/#2}\xspace}
}

\usepackage{latexsym}
\usepackage{amssymb}
\usepackage{amsmath} 
\usepackage{stmaryrd}
\usepackage{mathtools}
\usepackage{stackrel}

\usepackage{amsthm}
\newtheorem{theorem}{Theorem}[section]
\newtheorem{proposition}[theorem]{Proposition}
\newtheorem{lemma}[theorem]{Lemma}
\newtheorem{corollary}[theorem]{Corollary}
\theoremstyle{definition}
\theoremstyle{definition}
\theoremstyle{definition}\newtheorem{definition}[theorem]{Definition}
\theoremstyle{definition}
\theoremstyle{definition}\newtheorem{remark}[theorem]{Remark}
\theoremstyle{definition}

\newcommand{\sizeof}[1]{
  \left|#1\right|}

\newcommand{\ket}[1]{
    \ensuremath{\left|  #1 \right\rangle}\xspace}

\newcommand{\CX}{\ensuremath{\wedge X}\xspace}
\newcommand{\CNot}{\CX}

\ifx\numericids\undefined

\else

\fi

\newcommand{\zxcalculus}{\textsc{zx}-calculus\xspace}

\usepackage{dsfont}

\newcommand{\tket}{\ensuremath{\mathsf{t}|\mathsf{ket}\rangle}\xspace}

\usepackage{subcaption}
\usepackage{csvsimple}
\usepackage{multirow}
\usepackage{makecell}

\usepackage{hyperref}

%% file: tikzpreamble.tex
\usepackage{tikzit}  

\newcommand{\inlinetikzfig}[2][1.0]{
  \dimendef\grafheight=255\dimendef\grafvshift=254
  \setbox0 = \hbox{\scalebox{#1}{\input{#2.tikz}}}
  \grafheight=\the\ht0
  \grafvshift=-0.5\grafheight
  \advance\grafvshift by 0.5ex
  \raisebox{\grafvshift}{\input{#2.tikz}}
}

\newcommand{\inltf}[1]{\inlinetikzfig{#1}}

\pgfdeclarelayer{background} 
\pgfdeclarelayer{foreground}
\pgfdeclarelayer{edgelayer}
\pgfdeclarelayer{nodelayer}
\pgfsetlayers{background,edgelayer,nodelayer,main,foreground} 

\tikzstyle{halfsize}=[x=0.5cm, y=0.5cm]
\tikzstyle{normalsize}=[]
\tikzstyle{doublesize}=[]

\tikzstyle{(null)}=[]
\tikzstyle{plain}=[]

\usepackage{tikzzx}
\usepackage{tikzcircuits}
\usepackage[undirected]{tikzquanto} 
\usetikzlibrary{calc}

\tikzset{every picture/.style={line width=0.75pt}} 
\tikzset{directed edge/.style={postaction={decorate,decoration={markings,mark=at position 0.5 with {\arrow{>}}}}}}

%% file: abstract.tex
\begin{abstract}
  We give an overview of the circuit optimisation methods used by
  \tket, a compiler system for quantum software developed by Cambridge
  Quantum Computing Ltd.  We focus on a novel technique based around
  \emph{phase gadgets}, a family of multi-qubit quantum operations which
  occur naturally in a wide range of quantum circuits of practical
  interest.  The phase gadgets have a simple presentation in the
  \zxcalculus, which makes it easy to reason about them.  Taking
  advantage of this, we present an efficient method to translate the
  phase gadgets back to \CNot gates and single qubit operations
  suitable for execution on a quantum computer with significant
  reductions in gate count and circuit depth.  We demonstrate the
  effectiveness of these methods on a quantum chemistry benchmarking
  set based on variational circuits for ground state estimation of
  small molecules.
\end{abstract}

%% file: appendix.tex

\appendix

\section{Proof for Lemma~\ref{lem:1qb-cliff-rules}}\label{app:cliff}

\begin{proof}
A number of the rules follow from the ability to commute green vertices through $Z$ components of Pauli gadgets and red vertices through $X$ components using the spider fusion rule (S) and the colour-change rule (H) of the \zxcalculus.

\begin{equation}\label{eq:pauli-gphase}
{\inltf{PauliExpZga}}
\stackrel[(\ref{eq:pauli-exp-def})]{}{=} {\inltf{PauliExpZga1}}
\stackrel[(\text{S})]{}{=} {\inltf{PauliExpZga2}}
\stackrel[(\ref{eq:pauli-exp-def})]{}{=} {\inltf{PauliExpZgb}}
\end{equation}

\vspace{8mm}

\begin{equation}
\begin{aligned}
{\inltf{PauliExpXra}}
\stackrel[(\ref{eq:pauli-exp-def})]{}{=} {\inltf{PauliExpXra1}}
\stackrel[(\text{H})]{}{=} {\inltf{PauliExpXra2}} \\\\
\stackrel[(\text{S})]{}{=} {\inltf{PauliExpXra3}}
\stackrel[(\text{H}),(\ref{eq:pauli-exp-def})]{}{=} {\inltf{PauliExpXrb}}
\end{aligned}
\end{equation}

For the remaining $\pi$-phase properties, we will also need to use the $\pi$-copy/elimination rule (K1) and the phase-inversion rule (K2).

\begin{equation}\label{eq:pauli-pi}
\begin{aligned}
{\inltf{PauliExpXZa}}
\stackrel[(\ref{eq:pauli-exp-def})]{}{=} {\inltf{PauliExpXZa1}}
\stackrel[(\text{K1})]{}{=} {\inltf{PauliExpXZa2}}
\stackrel[(\text{S})]{}{=} {\inltf{PauliExpXZa3}} \\\\
\stackrel[(\text{K2})]{}{=} {\inltf{PauliExpXZa4}}
\stackrel[(\text{K1})]{}{=} {\inltf{PauliExpXZa5}}
\stackrel[(\text{S}), (\ref{eq:pauli-exp-def})]{}{=} {\inltf{PauliExpXZb}}
\end{aligned}
\end{equation}

The rest of the rules for passing single qubit Clifford gates through Pauli gadgets can be obtained straighforwardly using these, as in the following example.

\begin{equation}\label{eq:pauli-s}
\begin{aligned}
{\inltf{PauliExpSXa}}
\stackrel[(\ref{eq:pauli-exp-def})]{}{=} {\inltf{PauliExpSXa1}}
\stackrel[(\text{HD}), (\text{EU})]{}{=} {\inltf{PauliExpSXa2}}
\stackrel[(\ref{eq:pauli-pi})]{}{=} {\inltf{PauliExpSXa3}} \\\\
\stackrel[(\ref{eq:pauli-gphase})]{}{=} {\inltf{PauliExpSXa4}}
\stackrel[(\text{HD}), (\text{EU}), (\text{S}), (\text{I})]{}{=} {\inltf{PauliExpSXa5}}
\stackrel[(\ref{eq:pauli-exp-def})]{}{=} {\inltf{PauliExpSXb}}
\end{aligned}
\end{equation}
\end{proof}

\section{Proof for Lemma~\ref{lem:cnot-rules}}\label{app:cnot}

\begin{proof}
The control of a \CNot can commute through a $Z$ component of a Pauli gadget using just the spider fusion rule (S) of the \zxcalculus.

\begin{equation}
{\inltf{PauliExpCXZga}}
\stackrel[(\ref{eq:pauli-exp-def})]{}{=} {\inltf{PauliExpCXZga1}}
\stackrel[(\text{S})]{}{=} {\inltf{PauliExpCXZga2}}
\stackrel[(\ref{eq:pauli-exp-def})]{}{=} {\inltf{PauliExpCXZgb}}
\end{equation}

To prove the extension of Pauli gadget from a $X$ component, we remove Hadamard gates from the path with the colour-change rule (H) and introduce a pair of \CNot{}s using the identity (I) and Hopf (Hopf) rules. The rest follows from the bialgebra rule (B) and tidying up.

\begin{equation}\label{eq:cnot-Xg}
\begin{aligned}
{\inltf{PauliExpCXXga}}
\stackrel[(\ref{eq:pauli-exp-def})]{}{=} {\inltf{PauliExpCXXga1}}
\stackrel[(\text{H})]{}{=} {\inltf{PauliExpCXXga2}} \\\\
\stackrel[(\text{I}), (\text{Hopf})]{}{=} {\inltf{PauliExpCXXga3}}
\stackrel[(\text{S})]{}{=} {\inltf{PauliExpCXXga4}}
\stackrel[(\text{B})]{}{=} {\inltf{PauliExpCXXga5}} \\\\
\stackrel[(\text{H}), (\text{S})]{}{=} {\inltf{PauliExpCXXga6}}
\stackrel[(\ref{eq:pauli-exp-def})]{}{=} {\inltf{PauliExpCXXgb}}
\end{aligned}
\end{equation}

For the equivelent rule for $Y$, we spawn additional green phase vertices to allow us to introduce Hadamard gates via the Hadamard decomposition rule (HD), and reduce it to the $X$ case.

\begin{equation}
\begin{aligned}
{\inltf{PauliExpCXYga}}
\stackrel[(\ref{eq:pauli-exp-def})]{}{=} {\inltf{PauliExpCXYga1}}
\stackrel[(\text{I}), (\text{S})]{}{=} {\inltf{PauliExpCXYga2}} \\\\
\stackrel[(\text{HD})]{}{=} {\inltf{PauliExpCXYga3}}
\stackrel[(\ref{eq:pauli-exp-def})]{}{=} {\inltf{PauliExpCXYga4}}
\stackrel[(\ref{eq:cnot-Xg})]{}{=} {\inltf{PauliExpCXYga5}} \\\\
\stackrel[(\ref{lem:1qb-cliff-rules})]{}{=} {\inltf{PauliExpCXYga6}}
\stackrel[(\text{S})]{}{=} {\inltf{PauliExpCXYgb}}
\end{aligned}
\end{equation}

The remaining rules follow similarly.
\end{proof}
